\documentclass[sigconf, 10pt, nonacm]{acmart}

\AtBeginDocument{%
  \providecommand\BibTeX{{%
    \normalfont B\kern-0.5em{\scshape i\kern-0.25em b}\kern-0.8em\TeX}}}
    \usepackage{booktabs} 
 \usepackage{adjustbox}
\usepackage{xcolor}
\usepackage{epsfig,graphicx,parskip,setspace,tabularx,xspace}

\usepackage{amsmath}
\usepackage{caption}
\usepackage{subcaption}
\usepackage{booktabs}
\usepackage{todonotes}
\usepackage{cleveref}
\usepackage{soul}
\usepackage{siunitx}
\usepackage{algorithm}
\usepackage[noend]{algpseudocode}
\algrenewcommand{\algorithmiccomment}[1]{// #1}
\algrenewcommand{\ALG@beginalgorithmic}{\ttfamily\small}

\setcopyright{none}

\begin{document}

\title[A First Step Towards On-Device Monitoring of Body Sounds in the Wild]{A First Step Towards On-Device Monitoring\\of Body Sounds in the Wild}

\author{Shyam A. Tailor}
 \affiliation{%
  \institution{University of Oxford and University of Cambridge}
    }
\author{Jagmohan Chauhan}
 \affiliation{%
  \institution{University of Cambridge}
    }
    
    \author{Cecilia Mascolo}
 \affiliation{%
  \institution{University of Cambridge}
    }

 
\begin{abstract}
   Body sounds provide rich information about the state of the human body and can be useful in many medical applications.
   Auscultation, the practice of listening to  body sounds, has been used for centuries in respiratory and cardiac medicine to diagnose or track disease progression.
   To date, however, its use has been confined to clinical and highly controlled settings.
   Our work addresses this limitation: \emph{we devise a chest-mounted wearable for continuous monitoring of body sounds}, that leverages data processing algorithms that run \emph{on-device}.
   We concentrate on the detection of heart sounds to perform heart rate monitoring.
   To improve robustness to ambient noise and motion artefacts, our device uses an algorithm that explicitly segments the collected audio into the phases of the cardiac cycle.
   Our pilot study with 9 users demonstrates that it is possible to obtain heart rate estimates that are competitive with commercial heart rate monitors, with low enough power consumption for continuous use.


\end{abstract}


\maketitle


 \section{Introduction}\label{sec:introduction}


 Body sounds are a rich source of information about the health of an individual, and can be used for disease diagnosis or tracking. \emph{Auscultation}, the practice of listening to body sounds, has been used for centuries~\cite{tavelCardiacAuscultationGlorious2006} to diagnose cardiac, respiratory and digestive conditions.
 However, most existing research on auscultation is limited to user studies in a controlled and discrete (non-continuous) settings which limit the ability of the technique to detect diseases more widely and promptly.
 Continuous acoustic-based health monitoring has the potential to lead to affordable, continuous and scalable monitoring solutions, as microphones can be manufactured inexpensively and to fit wearable form factors.

 Recent works~\cite{rahmanBodyBeatMobileSystem2014, zhangpdvocal, larsonAccuratePrivacyPreserving2011} have explored the use of body sounds for medical diagnostics in non-clinical settings. BodyBeat~\cite{rahmanBodyBeatMobileSystem2014} captures sounds of food intake, breathing, laughter, and coughing to provide  information about the user's dietary behavior and respiratory physiology using a throat mounted microphone.
 PDVocal~\cite{zhangpdvocal} extracts non-speech body sounds from a user's phone usage in daily life to perform Parkinson's disease risk estimation.
 Other work using body sounds has focused on maintaining a healthy lifestyle~\cite{nandakumar2015contactless,  korpela2015evaluating}.
Complementing these works, we make a first step in constructing a continuous sensing wearable platform that can monitor body sounds accurately in realistic scenarios, where the user is performing natural activities in environments with significant ambient noise.

There are substantial challenges that must be considered to exploit audio collected from the body with a wearable device.
The device must be able to capture sounds of interest---meaning it must be sensitive to the appropriate frequencies, and be designed so it be placed near the source of the sound.
Additionally, the device must cope with high ambient noise levels and artefacts introduced by user activities, while running the inference on-device to \emph{maintain user privacy} and \emph{reduce power consumption}.
Keeping these challenges in mind, our overall aim \emph{is to open  doors to on-body acoustic-based health monitoring by offering the initial ingredients of a solution that allows on-device, continuous sound monitoring and analytics.}
This work focuses on heart rate monitoring as a motivating application.
However, the overarching goal of our research is to enable further research into continuously monitoring different vital signs, and the tracking of cardiac, respiratory and digestive diseases through the use of wearable devices exploiting an on-device analysis framework.

This paper makes  following contributions:
\begin{enumerate}
    \item We construct a novel device for collecting non-speech body sounds.
    The device is small enough to be worn under the user's clothes and its design explicitly considers recording sounds under challenging noise conditions.
    
    \item The collection of a new dataset, with 9 users, which assesses the impact of real world conditions on the device.
    The user study considers high ambient noise levels, and the effect of user activities, including motion.
    Prior work does not adequately assess these conditions.
    
    \item We propose an algorithm for extracting heart rate and heart rate variability (HRV).
    Results show that heart rate can be estimated accurately when participants are at rest and is competitive with commercially available chest-mounted heart rate trackers.
    The proposed algorithm runs efficiently on-device (continuously for 2 days) and is reasonably robust to noisy environments and  motion with median percentage error of  0.26 \(\pm\) 0.02\% under challenging noise conditions.

\end{enumerate}
\vspace{-2mm}
\section{Related Work}\label{sec:related}
\textbf{Applications of Non-Speech Body Sounds}:  Several existing works~\cite{rahmanBodyBeatMobileSystem2014, yataniBodyScopeWearableAcoustic2012, larsonAccuratePrivacyPreserving2011,   zhangpdvocal, nandakumar2015contactless} have explored monitoring body sounds using wearables.
Larson et al.~\cite{larsonAccuratePrivacyPreserving2011} detect coughs from audio captured by a mobile phone with an average true positive rate of 92\% and false positive rate of 0.5\%.  ApneaApp~\cite{nandakumar2015contactless} is a smartphone app which can detect sleep apnea issues in a contactless manner using respiratory sounds.
Yatani et al.~\cite{yataniBodyScopeWearableAcoustic2012} constructed a device to record sounds from the throat to perform human activity recognition.
These works highlight how body sounds are useful to track  health and well-being.
Our work differs in two ways.
Firstly, we focus on a continuous heart rate monitoring application rather than activity recognition or monitoring discrete respiration events; this is challenging especially in realistic settings.
Secondly, we focus on user privacy by performing all the  processing locally on-device.

\textbf{Heart Rate Monitoring}\label{sec:hrm}:
Different types of sensors such as photoplethysmography (PPG)~\cite{ phan2015smartwatch}, inertial~\cite{hernandez2015biowatch}, cameras ~\cite{kwon2012validation}, and wireless~\cite{adib2015smart} have been studied for heart rate monitoring.
Hernandez et al.~\cite{hernandez2015biowatch} showed that inertial measurement unit (IMU) sensors  can measure heart rate accurately from the wrists on some idle activities.
Adib et al.~\cite{adib2015smart} proposed doing heart rate estimation using WiFi signals.
Monitoring heart rate can be  highly inaccurate  with IMU and PPG sensors (especially from the wrist) as they are highly sensitive to human motion, and camera-based methods require the user to hold a camera in their hands and hence are impractical for continuous monitoring.
Wireless-based monitoring is impractical without wireless infrastructure. Due to these shortcomings, we see acoustics as an alternative for continuous heart rate monitoring.
Acoustic sensors are expected to be less impacted by motion than an IMU or PPG sensors but instead are sensitive to ambient noise.
We make initial steps to show that a device and algorithms can be  carefully designed to be resilient to noise or motion and perform robust heart rate estimations.

A recording of heart sounds is known as a phonocardiogram (PCG), and techniques for automated analysis have been investigated.
Applications include heart rate estimation or identifying heart abnormalities~\cite{springerRobustHeartRate2014, wangPhonocardiographicSignalAnalysis2007, rubinRecognizingAbnormalHeart2017, mohammadi2016wearable}.
However, these works are not usually analysed in the context of a limited computational or energy budget.
Many of these techniques require significant computational resources, which may be achieved using offloading from the wearable.
The downside with this approach, however, is that there are (justified) privacy concerns and issues with power consumption.
Our work aims to explicitly address these issues: we aim to perform continuous, on-device, heart rate monitoring in realistic settings.

\section{Wearable Device Design}\label{sec:design}

\textbf{Hardware}: We now give an overview of the hardware. 

\textit{Microphone Selection}: \label{sec:transducer_choice}
We used an off-the-shelf contact microphone~\cite{HIGHSENSITIVITYPIEZO}  that uses a piezoelectric transducer, rather than electret or condenser microphones, as they offer an excellent low-frequency response making them sensitive to the frequencies of interest: heart and lung sounds have substantial energy at near-infrasonic frequencies, 20-200 Hz.
These microphones pick up vibrations from the skin directly---necessitating the use of an elastic strap to hold the microphone against the skin---but making them less susceptible to ambient noise.
Other studies verified that this microphone is suitable for recording heart sounds~\cite{bifulcoMonitoringRespirationSeismocardiogram}.
We note that contact microphones perform sub-optimally with standard acoustic circuits: the signal may sound ``tinny'' as normal circuits are not designed for the piezoelectric elements, which are capacitive, and hence cause an impedance mismatch.
We will discuss the specifics of our acoustic circuitry when we describe the PCB implementation later in this section.

\textit{Microcontroller}: 
A Teensy 3.2 development board was used to control the wearable as it has low power consumption and has been successfully used for other audio datalogging projects~\cite{Tympan}.
Although we used the Teensy to collect the data, we do not foresee it being used in a commercial product due to reasons specified in section~\ref{sec:results}.

\textit{Printed Circuit Board Implementation}:  
The PCB design had 2 layers and  dimensions of 31x37mm.
Most board area is devoted to audio, but some ancillary functionality was included such as a microSD card slot and an IMU sensor (InvenSense MPU-9250) to collect inertial data. 

Connecting the microphone directly to the analog-digital converter (ADC) would yield poor results due to a poor impedance match which would high-pass the signal.
The microphone output would also not use the full ADC range, resulting in lower resolution recordings than otherwise achievable.
It is necessary to amplify the signal while preserving the low-frequency components.
A buffer circuit was built using a single operation amplifier stage to improve the impedance match.
Piezo voltage spikes were handled using transient voltage spike diodes. The buffered signal was fed into passive RC high-pass and low-pass filter stages and amplified using non-inverting amplifiers.
The amplified signal was then connected to the Teensy's ADC.
The final PCB design used a low-power audio operational amplifier~\cite{OPA1692SoundPlusLowPower} designed for wireless microphones.
An enclosure was designed using Autodesk Fusion 360 and 3D printed; the enclosure was mounted onto an elastic strap to be worn around the torso.  A button and LED was added to allow participants to start the recording and monitor the wearable status.

\textbf{Software}: A datalogging program was written using C++.
The program continuously collected audio and inertial data and wrote it to a microSD card.

\textbf{Placement}: Four possible placements, often evaluated in clinical settings, were considered:  (1) bottom of sternum, (2) on the back behind the heart, (3) bottom of ribcage (offset towards  heart) and (4) at the top of the chest (offset towards  heart).
The last placement was chosen to align with the bronchus, which was hypothesised to allow better transmission of respiratory sounds.
Placement 1 was unstable due to the surrounding curvature on the body, which has a wide inter-person variance.
Placement 2 is also problematic: shoulder blade movement displaces the device from the skin.
Placement 3 and 4 were evaluated by one of the authors.
The top of the chest was eventually selected after assessing the amplitude of the sounds recorded while performing 7 different activities, including sitting quietly, deeply breathing, drinking, coughing, sniffing, throat clearing, and talking.
\section{Heart Rate Estimation Algorithm}\label{sec:algos}
Our proposed algorithm uses the sounds present in the cardiac cycle.
We first discuss the cardiac cycle and the associated sounds, before proceeding to describe heart rate and HRV estimation algorithms.

\textbf{Cardiac Cycle}: \label{sec:cardiac_cycle} The heart consists of 4 chambers: 2 atria, and 2 ventricles~\cite{gershMayoClinicHeart2000}.
The atria force blood into the ventricles, which then pump the blood into the arteries.
There are 2 pairs of valves in the heart. 
Atrioventricular (AV) valves allow blood to flow from the atria to the ventricles.
Semilunar (SL) valves allow blood to flow from the ventricles into the arteries.
The cardiac cycle proceeds as follows: (1) AV and SL valves are shut; blood flows  into the atria from the veins, (2a) AV valves open, and blood starts to flow from the atria into the ventricles, (2b) Atria contract to force blood into the ventricles,  (3) AV valves shut, (4) Ventricles contract, forcing the SL valves to open and let blood into the arteries.
SL valves close to prevent blood flowing back into the ventricles.

\begin{figure}[h]
    \includegraphics[width=0.5\textwidth]{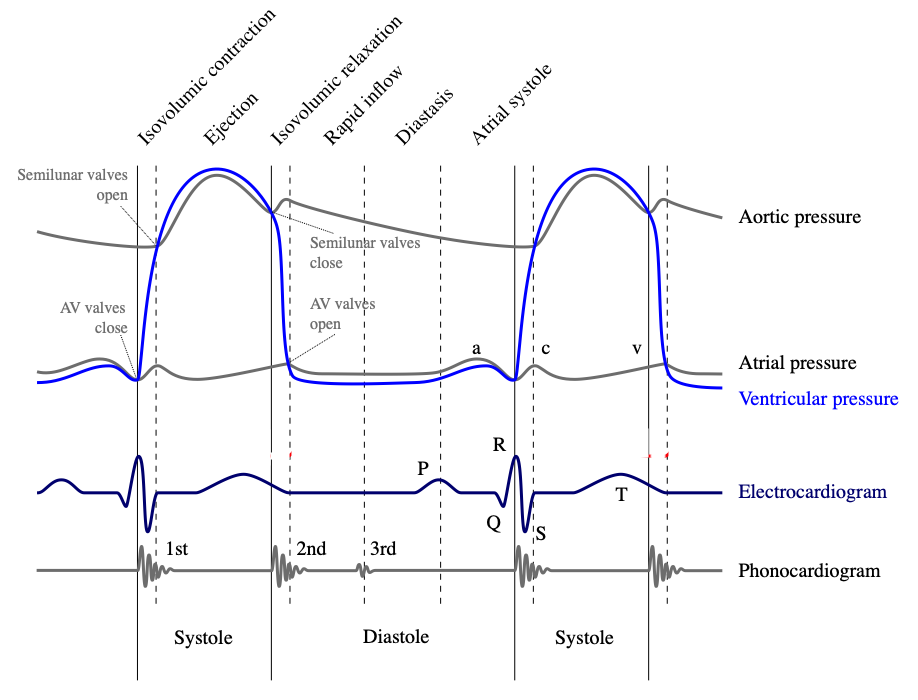}
        \caption{Heart sounds in cardiac cycle~\cite{xavaxEnglishWiggersDiagram2012})}
    \label{fig:wiggers}
\end{figure}

The first two phases are known as the diastole, where the heart is relaxed.
The last two phases form the systole period, where the heart is beating. From a healthy heart, two distinct sounds are heard  corresponding to valve closures.
The first sound, \emph{S\(_1\)}, referred to as ``lub'', occurs at the start of systole when the AV valves shut.
The second sound, \emph{S\(_2\)} (``dub''), corresponds to  SL valves shutting, and is usually quieter than the S\(_1\) as shown in Figure \ref{fig:wiggers}.
P, Q, R, S, and T are points on an aligned electrocardiogram (ECG) waveform.

\textbf{Heart Rate Estimation Using Segmentation}: Springer et al.~\cite{springerLogisticRegressionHSMMBasedHeart2016} proposed a state-of-the-art algorithm for PCG segmentation, which used a hidden \emph{semi}-markov model (HSMM) to model transitions between 4 states: S\(_1\), systole, S\(_2\) and diastole~\cite{springerLogisticRegressionHSMMBasedHeart2016}.
Using an HSMM allows for the state residence time to be modelled, as the transition probabilities are dependent on the time that has elapsed since entry to the current state.
An ordinary HMM models residence distributions as geometric, which is a poor approximation for this problem.
The emission probabilities for each state is modelled using logistic regression trained from 4 features extracted from the PCG.
Each feature is a type of envelope extracted from the signal and labelled as one of the 4 states using reference ECG data.
The results on our collected data were extremely poor when using this algorithm as the original features proposed are not noise-robust: envelope-based features are disrupted by transient changes in signal amplitude.
Another issue is that these features are computationally expensive.
Hence, we propose several modifications to the original algorithm. 

\begin{figure}
    \centering
    \begin{subfigure}[t]{.45\textwidth}
        \centering
        \includegraphics[width=\textwidth]{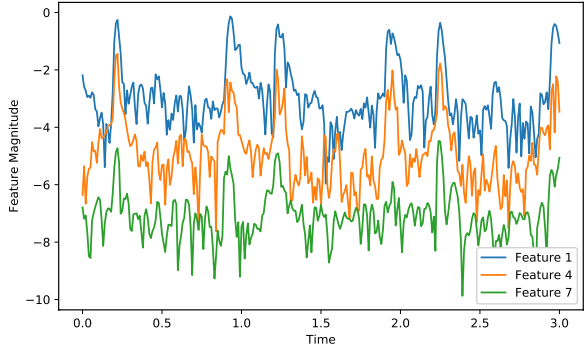}
        \caption{Still}
    \end{subfigure}
    \begin{subfigure}[t]{.45\textwidth}
        \centering
        \includegraphics[width=\textwidth]{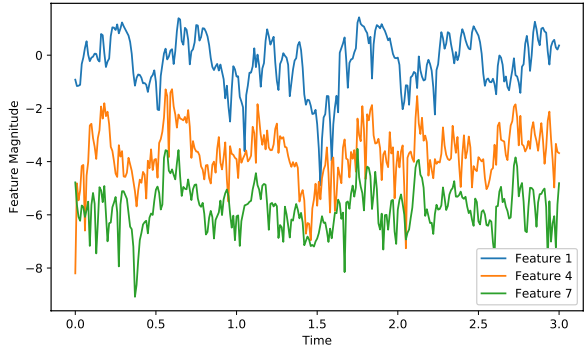}
        \caption{Walking}
    \end{subfigure}
    \caption{
    Plot of spectral features used for segmentation against time.
    Higher-indexed features correspond to higher frequencies.
    }
    \label{fig:segment_features}
\end{figure}

\textit{Adaptations for Continuous Monitoring}: Our algorithm is described in Algorithm \ref{alg:heart_segment}. We used log-magnitudes of Short-Time Fourier Transform (STFT) coefficients as features: it was observed that different activities introduced power into different parts of the spectrum.
In principle, a classifier could learn to ignore spikes in coefficients that correspond to known noise profiles---so long as the rest of the coefficients are consistent with a heart sound.
We used audio sampling rate of 500Hz to enable frequencies below 200Hz to be distinguished.
The STFT was performed on 16 samples (Hann windowing, hop length of 5 samples) yielding 100 (500/5) \(\times\) 9 (16/2 + 1) features per second.
A plot of the features against time is reported in Figure~\ref{fig:segment_features}.
When the user is stationary, consistent spikes in all features can be observed, corresponding to heart sounds.
Determining the location of the heart sounds from walking data is difficult, however it is not impossible.
As explained, the noise introduced by walking primarily affects low frequencies (sub-100Hz), and hence a classifier could learn to handle such noises.

\begin{algorithm} [t]
\begin{algorithmic}
\Function{SegmentHeartCycle}{$\mathbf{x}$, classifier}
    \State\Comment{log-magnitudes of STFT coefficients}
    \State features = \Call{ExtractFeatures}{$\mathbf{x}$}
    \State probs = \Call{estimateEmissionProbs}{classifier, features}
    \State\Comment{Viterbi decoding for a HSMM}
    \State \Return \Call{HsmmDecode}{4, Len(probs), probs, $\ldots$} 
\EndFunction
\end{algorithmic}
\caption{Heart rate segmentation.}\label{alg:heart_segment}
\end{algorithm}

We also used Random Forests (RF) instead of  logistic regression (LR) as a RF can learn non-linear decision boundaries, while admitting efficient inference on a microcontroller.
Non-linear decision boundaries are needed to allow the classifier to cope with noisy measurements.
A RF (10 trees, maximum depth 8) was trained to predict the presence of either type of heart sound, reducing it to binary classification: at time \(t\) it was assumed that the emission probability for the S\(_1\) (systole) and S\(_2\) (diastole)  states was the same.
This optimisation was made as the two heart sounds are difficult to distinguish, as they differ mainly in duration.
By tying the emission probabilities, the inference of the most likely path through the states relies on duration of the states observed.
This is relatively easy as the systole and diastole states have notably different durations: 0.128 \(\pm\) 0.062 and 0.356 \(\pm\) 0.121 seconds for the two states were observed in our dataset.

\textit{Estimating Heart Rate and Variability}: The labelling generated by Algorithm~\ref{alg:heart_segment} is post-processed by Algorithm \ref{alg:segment_derived}, which describes steps to obtain the heart rate and HRV.
The intervals between heartbeats are found from the labels; intervals are the differences between the \emph{start} of two S\(_1\) states.
Kalman filters were used to estimate heart rate from the intervals.
For accurate HRV estimation outliers were rejected using a standard technique~\cite{kemperHeartRateVariability2007a}.

\begin{algorithm}
\begin{algorithmic}
\Function{EstimateHeartRate}{labels}
    \State deltas = \Call{FindTimesBetweenBeats}{labels}
    \State bpm = 60 / (deltas / labelSampleRate)
    \State bpm = \Call{KalmanFilter}{bpm}
    \State \Return bpm
\EndFunction
\Function{EstimateVariability}{labels}
    \State deltas = \Call{FindTimesBetweenBeats}{labels}
    \State retainedDeltas = List()
    \For{i in 1$\ldots$Len(deltas) - 1}
        \State\Comment{Find mean of 4 previous inter-beat times}
        \State window = deltas[\Call{Max}{i - 4, 0}: i]
        \State m = \Call{Mean}{window}
        \State\Comment{Reject if more than 30\% away from local mean}
        \If{m $\times$ 0.7  $\leq$ deltas[i] $\leq$ m $\times$ 1.3}
            \State retainedDeltas.Append(deltas[i])
        \EndIf
    \EndFor
    \State \Return \Call{StandardDeviation}{retainedDeltas}
\EndFunction
\end{algorithmic}
\caption{
HR and HRV estimation.
}\label{alg:segment_derived}
\end{algorithm}
\vspace{-3mm}
\section{User Study}\label{sec:collection}
\begin{figure}
    \centering
    \includegraphics[width=0.2\textwidth]{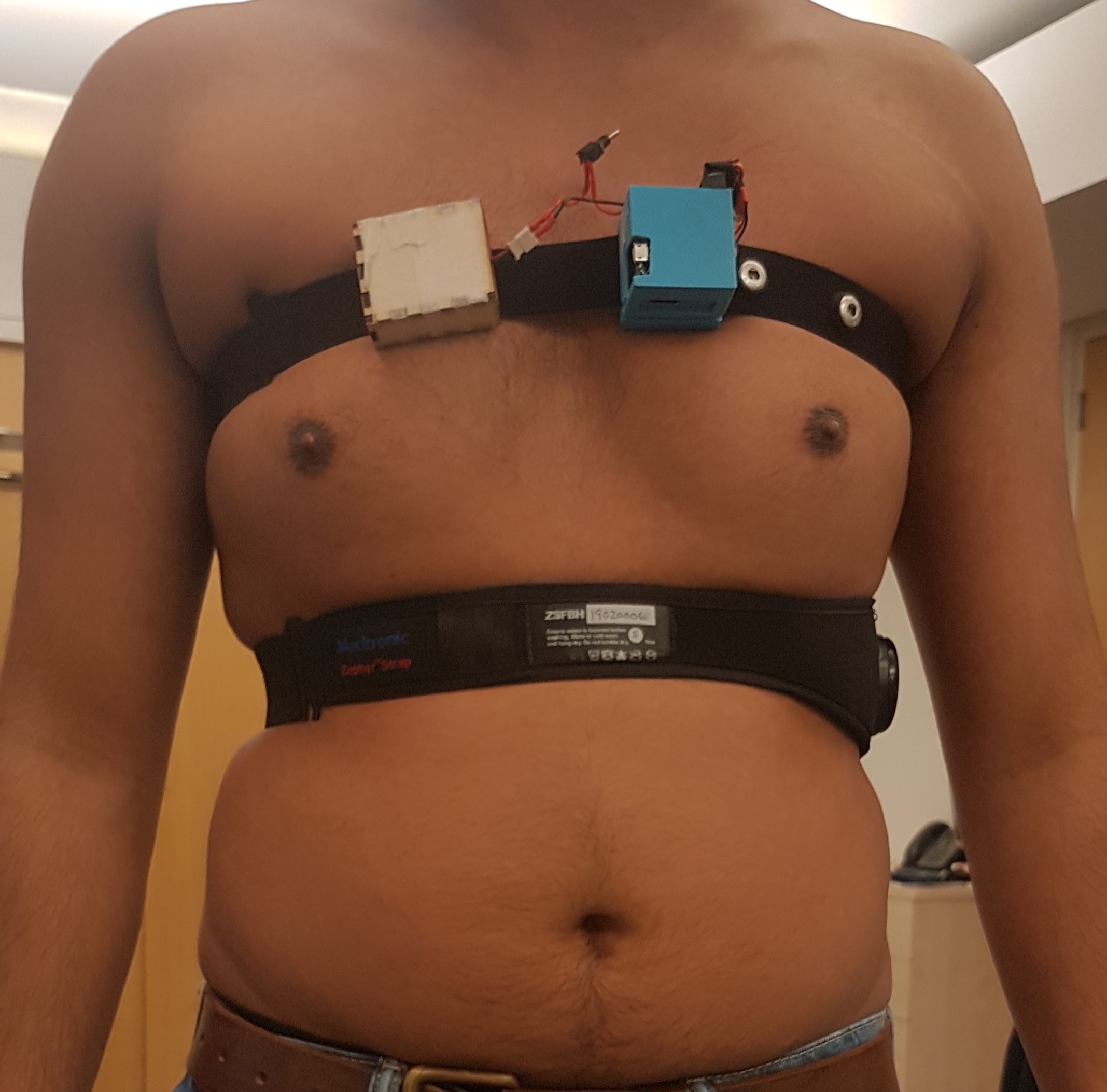}
    \caption{Devices: Ours (top),  ground truth (bottom).}
    \label{fig:wearing_devices}
\end{figure}

Nine participants (3 female, 6 male, all healthy) performed the data collection after we obtained ethics approval.
Participants wore the device (Figure \ref{fig:wearing_devices}) and  perform several activities under 3 noise regimes: silence, music placed 1 meter from the participant (average loudness of ~63dB) and background noise from a coffee shop placed 1 meter from the participant (average loudness of ~42dB).
Participants did the following activities: breathing normally and deeply, coughing, clearing throat and sniffing  10 times,  swallowing 5 times, drinking water, reading a news article, and walking and jogging for 5 minutes.
A Zephyr Bioharness 3 was also worn to provide ground truth ECG data.
Inertial data was used to synchronize recordings between the two devices and detect  motion.

\section{Results}\label{sec:results}

\textbf{Accuracy}:  The results reported are an average of 10 runs using leave-one-person-out cross validation.

\textit{Processing ECG based Ground Truth Data}: 
Raw ECG data was processed using BioSPPY~\cite{WelcomeBioSPPyBioSPPy} and Neurokit~\cite{PythonToolboxStatistics2019} libraries to obtain heart rate estimates from the ground truth data.
Using the raw signal derived from the inter-beat-intervals yields noisy results because of HRV. To compensate, we applied exponential smoothing to the raw signal i.e.~\(s(t) = \alpha x(t) + (1 - \alpha)s(t-1)\).
\(\alpha = 0.075\) was chosen as a compromise between rejecting high-frequency noise while still preserving local trends.
We followed the same procedure as outlined in Algorithm~\ref{alg:segment_derived} for HRV estimation.

\textit{Heart Rate Estimation}: 
The  heart rate estimates were compared to the ground truth every 2 seconds and are shown in Table~\ref{tab:seg_hr_res}
The results shows that accuracy when resting is competitive with commercially available chest-mounted heart rate trackers: one study indicated a popular device had a mean percentage error of 0.8\%~\cite{gillinovVariableAccuracyWearable2017}.
We obtained a mean percentage error of approximately 3.34\% in the worst case, but  median percentage error was below 0.33\% for each scenario.
Different noise regimes have minimal impact, demonstrating our approach's noise robustness.

Reasonable heart rate estimates can also be obtained when doing other activities: even while walking, the median percentage error was 7.35\%.
This result confirms our hypothesis that the classifier can learn to classify heart beats from the spectral coefficients, even when some noise is introduced.
However, further noise, as observed with running, causes the algorithm to completely degrade.
Performance also degrades during speech; this occurs for both genders.
We believe that this is due to the presence of low frequency respiratory noise rather than the vocal frequencies.

\begin{table}[t]
\centering
\resizebox{.5\textwidth}{!}{\begin{tabular}{@{}llrrrr@{}}
\toprule
\textbf{Activity} & \parbox[c]{2cm}{\raggedright\textbf{Noise Regime}} & \parbox[c]{2cm}{\raggedleft\textbf{Median Absolute Error / BPM}} & \parbox[c]{2cm}{\raggedleft\textbf{Mean Absolute Error / BPM}} & \parbox[c]{2.5cm}{\raggedleft\textbf{Median Percentage Error}} & \parbox[c]{2.5cm}{\raggedleft\textbf{Mean Percentage Error}} \\ \midrule
Rest & Silence & 0.23 $\pm$ 0.01 & 1.18 $\pm$ 0.08 & 0.33 $\pm$ 0.02 & 1.49 $\pm$ 0.09 \\
Rest & Music & 0.20 $\pm$ 0.01 & 2.91 $\pm$ 0.11 & 0.26 $\pm$ 0.02 & 3.34 $\pm$ 0.13 \\
Rest & Conversation & 0.14 $\pm$ 0.04 & 2.54 $\pm$ 0.12 & 0.22 $\pm$ 0.07 & 2.77 $\pm$ 0.13 \\
Deep Breathing & Silence & 5.18 $\pm$ 0.27 & 7.15 $\pm$ 0.11 & 6.04 $\pm$ 0.32 & 7.80 $\pm$ 0.12 \\
Deep Breathing & Music & 11.84 $\pm$ 0.43 & 14.20 $\pm$ 0.21 & 13.91 $\pm$ 0.56 & 15.07 $\pm$ 0.24 \\
Deep Breathing & Conversation & 8.47 $\pm$ 0.74 & 12.29 $\pm$ 0.30 & 9.94 $\pm$ 0.82 & 13.25 $\pm$ 0.34 \\
Coughing & Silence & 13.75 $\pm$ 1.45 & 13.61 $\pm$ 0.61 & 14.24 $\pm$ 1.38 & 14.33 $\pm$ 0.67 \\
Coughing & Music & 5.81 $\pm$ 0.40 & 8.59 $\pm$ 0.23 & 6.98 $\pm$ 0.49 & 9.53 $\pm$ 0.29 \\
Coughing & Conversation & 5.32 $\pm$ 0.32 & 8.73 $\pm$ 0.19 & 6.75 $\pm$ 0.57 & 9.57 $\pm$ 0.23 \\
Clearing Throat & Silence & 5.40 $\pm$ 0.57 & 7.90 $\pm$ 0.35 & 6.71 $\pm$ 0.85 & 8.96 $\pm$ 0.41 \\
Clearing Throat & Music & 4.14 $\pm$ 0.50 & 7.27 $\pm$ 0.33 & 5.70 $\pm$ 0.67 & 8.68 $\pm$ 0.49 \\
Clearing Throat & Conversation & 1.40 $\pm$ 0.18 & 4.78 $\pm$ 0.36 & 1.91 $\pm$ 0.19 & 5.31 $\pm$ 0.40 \\
Swallowing & Silence & 2.51 $\pm$ 0.78 & 5.65 $\pm$ 0.27 & 2.93 $\pm$ 0.90 & 6.28 $\pm$ 0.32 \\
Swallowing & Music & 3.24 $\pm$ 0.86 & 5.42 $\pm$ 0.31 & 3.98 $\pm$ 0.81 & 6.19 $\pm$ 0.35 \\
Swallowing & Conversation & 3.09 $\pm$ 0.53 & 8.11 $\pm$ 0.18 & 4.33 $\pm$ 0.76 & 9.46 $\pm$ 0.24 \\
Drinking & Silence & 7.07 $\pm$ 0.27 & 7.97 $\pm$ 0.30 & 8.23 $\pm$ 0.38 & 9.18 $\pm$ 0.34 \\
Sniffing & Silence & 2.89 $\pm$ 0.41 & 4.63 $\pm$ 0.29 & 3.63 $\pm$ 0.34 & 5.73 $\pm$ 0.44 \\
Sniffing & Music & 1.02 $\pm$ 0.22 & 3.85 $\pm$ 0.39 & 1.45 $\pm$ 0.35 & 4.79 $\pm$ 0.55 \\
Sniffing & Conversation & 1.18 $\pm$ 0.10 & 3.77 $\pm$ 0.22 & 1.64 $\pm$ 0.11 & 4.31 $\pm$ 0.26 \\
Speech & Silence & 7.55 $\pm$ 0.41 & 10.81 $\pm$ 0.26 & 11.11 $\pm$ 0.56 & 13.85 $\pm$ 0.37 \\
Walking & Silence & 6.11 $\pm$ 0.10 & 8.56 $\pm$ 0.09 & 7.35 $\pm$ 0.17 & 9.29 $\pm$ 0.11 \\
Running & Silence & 37.61 $\pm$ 0.40 & 38.99 $\pm$ 0.25 & 31.62 $\pm$ 0.31 & 30.64 $\pm$ 0.20 \\ \bottomrule
\end{tabular}}
\caption{
Accuracy of segmentation-based  algorithm.
}
\label{tab:seg_hr_res}
\vspace{-5mm}
\end{table}

\textit{Segmentation Accuracy}: 
\begin{table}[t]
\centering
\resizebox{.5\textwidth}{!}{\begin{tabular}{@{}llrrr@{}}
\toprule
\textbf{Activity} & \parbox[c]{2cm}{\raggedright\textbf{Noise Regime}} & \parbox[c]{2cm}{\raggedleft\textbf{Precision}} & \parbox[c]{2cm}{\raggedleft\textbf{Recall}} & \parbox[c]{2.5cm}{\raggedleft\textbf{F1}} \\ \midrule
Rest & Silence & 0.976 $\pm$ 0.001 & 0.960 $\pm$ 0.003 & 0.968 $\pm$ 0.002 \\
Rest & Music & 0.937 $\pm$ 0.004 & 0.893 $\pm$ 0.004 & 0.915 $\pm$ 0.004 \\
Rest & Conversation & 0.963 $\pm$ 0.004 & 0.934 $\pm$ 0.006 & 0.948 $\pm$ 0.005 \\
Deep Breathing & Silence & 0.864 $\pm$ 0.004 & 0.772 $\pm$ 0.004 & 0.815 $\pm$ 0.004 \\
Deep Breathing & Music & 0.776 $\pm$ 0.007 & 0.637 $\pm$ 0.008 & 0.699 $\pm$ 0.007 \\
Deep Breathing & Conversation & 0.819 $\pm$ 0.004 & 0.703 $\pm$ 0.005 & 0.757 $\pm$ 0.005 \\
Coughing & Silence & 0.693 $\pm$ 0.011 & 0.568 $\pm$ 0.010 & 0.624 $\pm$ 0.011 \\
Coughing & Music & 0.768 $\pm$ 0.009 & 0.682 $\pm$ 0.011 & 0.722 $\pm$ 0.010 \\
Coughing & Conversation & 0.749 $\pm$ 0.011 & 0.655 $\pm$ 0.011 & 0.699 $\pm$ 0.011 \\
Clearing Throat & Silence & 0.725 $\pm$ 0.012 & 0.658 $\pm$ 0.011 & 0.690 $\pm$ 0.011 \\
Clearing Throat & Music & 0.820 $\pm$ 0.013 & 0.765 $\pm$ 0.012 & 0.791 $\pm$ 0.013 \\
Clearing Throat & Conversation & 0.783 $\pm$ 0.011 & 0.736 $\pm$ 0.009 & 0.759 $\pm$ 0.010 \\
Swallowing & Silence & 0.902 $\pm$ 0.015 & 0.865 $\pm$ 0.021 & 0.883 $\pm$ 0.018 \\
Swallowing & Music & 0.866 $\pm$ 0.006 & 0.790 $\pm$ 0.010 & 0.827 $\pm$ 0.008 \\
Swallowing & Conversation & 0.829 $\pm$ 0.013 & 0.752 $\pm$ 0.013 & 0.789 $\pm$ 0.013 \\
Drinking & Silence & 0.801 $\pm$ 0.007 & 0.717 $\pm$ 0.005 & 0.756 $\pm$ 0.006 \\
Sniffing & Silence & 0.832 $\pm$ 0.005 & 0.797 $\pm$ 0.006 & 0.814 $\pm$ 0.005 \\
Sniffing & Music & 0.822 $\pm$ 0.008 & 0.790 $\pm$ 0.008 & 0.805 $\pm$ 0.008 \\
Sniffing & Conversation & 0.818 $\pm$ 0.008 & 0.785 $\pm$ 0.008 & 0.801 $\pm$ 0.008 \\
Speech & Silence & 0.652 $\pm$ 0.008 & 0.599 $\pm$ 0.007 & 0.624 $\pm$ 0.008 \\
Walking & Silence & 0.548 $\pm$ 0.005 & 0.494 $\pm$ 0.005 & 0.519 $\pm$ 0.005 \\
Running & Silence & 0.456 $\pm$ 0.004 & 0.309 $\pm$ 0.003 & 0.368 $\pm$ 0.003 \\ \bottomrule
\end{tabular}}
\caption{
Accuracy of S\(_1\) heart sound localisation.
}
\label{tab:seg_s1_acc}
\end{table}
The segmentation quality is assessed by evaluating the precision and recall of the predicted S\(_1\) states, relative to the ground truth ECG.
If the \emph{start} of the S\(_1\) state occurs within 100ms of an R-Peak in the corresponding ECG signal, then the segmentation algorithm is deemed to have correctly predicted the S\(_1\) state.
The results are reported in Table~\ref{tab:seg_s1_acc} and are comparable to Springer et al.~\cite{springerLogisticRegressionHSMMBasedHeart2016} (F1 0.956) when users are at rest---despite their data coming from a clinical digital stethoscope.
We observe a larger than expected difference between precision and recall for deep breathing: this is likely due to a medical phenomenon where the heart sounds can split during breathing, making distinguishing them more difficult.

Despite the poor results for segmentation during walking, the heart rate estimates for walking are reasonably accurate.
This is because of the emission probability estimates for footsteps being high enough that the Viterbi algorithm takes them for a heart sound, as there is a plausible time difference between the footstep and an actual heart sound.
This leads to the true S\(_1\) sound being reported as the S\(_2\) sound, and the S\(_2\) sound not being identified, as it has a lower energy than the S\(_1\) sound.

\textit{Heart Rate Variability}: 
\begin{table}[t]
\centering
\resizebox{.5\textwidth}{!}{\begin{tabular}{@{}llrrrr@{}}
\toprule
\textbf{Activity} & \parbox[c]{2cm}{\raggedright\textbf{Noise Regime}} & \parbox[c]{2cm}{\raggedleft\textbf{Median Absolute Error / ms}} & \parbox[c]{2cm}{\raggedleft\textbf{Mean Absolute Error / ms}} & \parbox[c]{2.5cm}{\raggedleft\textbf{Median Percentage Error}} & \parbox[c]{2.5cm}{\raggedleft\textbf{Mean Percentage Error}} \\ \midrule
Rest & Silence & 4.01 $\pm$ 0.79 & 8.28 $\pm$ 1.30 & 6.14 $\pm$ 1.19 & 14.51 $\pm$ 2.66 \\
Rest & Music & 7.50 $\pm$ 6.10 & 19.18 $\pm$ 2.77 & 14.74 $\pm$ 9.37 & 39.25 $\pm$ 10.13 \\
Rest & Conversation & 7.78 $\pm$ 3.27 & 9.81 $\pm$ 1.43 & 13.26 $\pm$ 4.75 & 20.38 $\pm$ 3.10 \\ \bottomrule
\end{tabular}}
\caption{
Accuracy of  HRV  estimates.
}
\label{tab:seg_hrv}
\end{table}
Table~\ref{tab:seg_hrv} only gives the results for HRV estimation when  participants were at rest  as  HRV requires accurate segmentation.
To the best of the authors' knowledge, no empirical survey has assessed HRV accuracy for commercially available heart rate monitors which makes it difficult to compare.
Values reported for HRV in healthy adults had an inter-quartile range of approximately 30ms~\cite{keetShorttermHeartRate2013}; the accuracy obtained by our approach is hence likely sufficient for indicative readings.

\textbf{Power Consumption and Latency}\label{sec:power_eval_heart}:
An STMicroelectronics Nucleo L496ZG-P board  (ARM Cortex-M4F, 80MHz, 320KB of SRAM) was used to run latency and power consumption experiments as it is a realistic, low power, microcontroller that could feasibly be used in a commercial product.
We used the Teensy to reduce development time, but we do not expect a commercial device to be based upon it.
Another benefit of using the STM board is that it allows accurate power measurements to be taken.

\textit{Latency}: To measure latency, the clock cycle register is used.
It took 3ms to extract 1 second of audio features,  approximately 1.2ms to calculate 1 second of emission probabilities and 695.3 ms to run Viterbi algorithm.
Therefore, assuming overheads, the segmentation algorithm requires approximately 700ms of computation for 1s of audio.

\textit{Power Consumption}: The biggest consumer of power was the microcontroller: 4.5mA during run-mode and \SI{276}{\micro\ampere} during sleep (with ADC sampling enabled), at 3.3V.
As a contact microphone is a passive sensor it does not need external power to make readings, and amplification uses negligible power.
Assuming a 3.7V 200mAh lithium-polymer battery and a safety margin of 0.7, the battery life is approximately \emph{48 hours}.

\section{Conclusion}\label{sec:discussion}
Body sounds are an excellent source of information for understanding the state of the human body and are being increasingly explored in the area of digital health.
We designed a wearable device to assess the viability of monitoring heart rate continuously using heart sounds.
This work has explicitly considered the difficult challenges associated with our goal: \emph{noisy measurements} and \emph{limited computational and energy budgets}.
Our work makes an important step towards enabling heart sounds to be monitored continuously in non-clinical conditions, and we believe that similar approaches can be adopted for other types of body sounds.
\vspace{-3mm}
\begin{acks}
The authors would like to thank Brian Jones along with the other members of the Mobile Systems Group.
This work was supported by ERC Project 833296 (EAR).
\end{acks}
\vspace{-3mm}
\bibliographystyle{ACM-Reference-Format}
\bibliography{acm}
\end{document}